\documentclass{ws-ijmpa}

\begin{document}

\markboth{H. Ray}
{Current Status of the MiniBooNE Experiment}

%
\catchline{}{}{}{}{}
%

\title{Current Status of the MiniBooNE Experiment }

\author{\footnotesize H. Ray for the MiniBooNE collaboration~\cite{MiniBooNE}\footnote{
current address : Fermi National Laboratory,
          MiniBooNE c/o M.S. 309,
          Batavia, IL 60510, U.S.A} }

\address{Los Alamos National Laboratory \\ 
Los Alamos, New Mexico  87545, U.S.A }

\maketitle

\begin{abstract}
MiniBooNE is an experiment designed to refute or confirm the 
LSND $\overline{\nu}_{\mu} \rightarrow \overline{\nu}_e$ oscillation result.  MiniBooNE will look for oscillations of 
$\nu_{\mu} \rightarrow \nu_{e}$ in a closed-box appearance analysis.
MiniBooNE began collecting data in 2002, and is expected to continue data taking through 2005.  Current 
MiniBooNE results are presented.
\end{abstract}

\section{Motivation for MiniBooNE}
Neutrino oscillations have been observed at three different mass scales :   $\triangle \mathrm{m_{solar}}^2$ $\approx$
8x$10^{-5} \mathrm{eV}^2$, $\triangle \mathrm{m_{atm}}^2$ $\approx$ 2x$10^{-3} \mathrm{eV}^2$, and 
$\triangle \mathrm{m_{LSND}}^2$ $\approx 0.1 \rightarrow 10 \ \mathrm{eV}^2$.  
In the Standard Model there are three active neutrinos, and a summation law exists such that  $\triangle \mathrm{m_{12}}^2$ + 
$\triangle \mathrm{m_{23}}^2$  = $\triangle \mathrm{m_{13}}^2$.  
The three mass scales of observed oscillations do not follow this law.  Therefore, one of the results is faulty or there exists physics beyond the Standard Model.  
Solar and atmospheric oscillations have been observed and confirmed by several experiments.  
MiniBooNE will make a definitive test of the LSND~\cite{LSND} signal.

\section{The MiniBooNE Detector}

MiniBooNE is a fixed target experiment located at Fermi National Laboratory.  
The Fermilab Booster provides MiniBooNE with an 8 GeV proton beam.
Protons are directed onto a beryllium target located inside a magnetic focusing horn. 
Approximately 3x$10^{12}$ protons per 1.6 $\mu s$ pulse arrive at the MiniBooNE target at a rate of 5 Hz.
The interaction of the protons with the target produces a stream of charged mesons.  
Magnetically focusing the mesons increases the neutrino flux by a factor of 6.  
After focusing, the mesons
travel into a 50 meter decay region where they decay in flight to produce the $\nu_{\mu}$ beam.  
The $\nu_{\mu}$ travel through 490 meters of a dirt absorber before entering the 
MiniBooNE detector.  
The energy of these neutrinos is $\approx$ 700 MeV, providing an L/E of  
$\approx$ 0.8 m/MeV (compared to LSND L/E of $\approx$1 m/MeV).  

The MiniBooNE detector is a 12 meter diameter sphere, filled with undoped mineral oil.  The tank consists of two 
regions : an inner light-tight region, lined with 1280 PMTs to provide 10\% photocathode coverage, and an outer 
optically isolated veto-region containing 240 PMTs.  
Different beam energy, beam duty cycle, and oil give MiniBooNE drastically different 
systematic errors than those found at LSND.

\section{Basic Analysis Components}
There are three main components to all of the analyses at MiniBooNE:
flux prediction, calibration of the data, and particle identification.

The neutrino beam which enters the MiniBooNE tank contains $\nu_{\mu}$
 and some intrinsic $\nu_e$.  The intrinsic $\nu_e$ is thought to be 
$\approx$ 0.5\% of the $\nu_{\mu}$ flux 
and is comparable to the amount of $\nu_e$ expected from a positive oscillation signal.  
MiniBooNE relies on data from the E910~\cite{E910} and HARP~\cite{HARP}
 experiments to aid in understanding beam flux and cross sections.  These two experiments directed a proton beam in the MiniBooNE energy range (8-12 GeV) onto a beryllium target.  
The \emph{in situ} kaon flux at MiniBooNE is measured using a scintillating fiber tracker located in a small 
beam-pipe at an angle of 6 degrees to the decay region.  
Cross sections are found using MiniBooNE data as well as data from other experiments.

Seven optically isolated scintillating cubes 
located throughout the detector, combined with a muon tracker located directly above the tank, identify stopping muons and allow for identification 
of a sample of Michel electrons of known position for energy calibration.  
The high end of the Michel electron energy spectrum is used to fix the detector energy scale.  
MiniBooNE currently has a 14.8\% energy reconstruction at 50 MeV.  

MiniBooNE has yet to settle on a final particle ID algorithm for the oscillation analysis.  
We are exploring the use of neural nets and boosting algorithms.~\cite{Roe04}

\section{Physics Updates}

The goal of the MiniBooNE oscillation $\nu_e$ appearance analysis is to explore all of the LSND 90\% 
confidence region to the 4-5 sigma level.  With our current estimates of neutrino fluxes and reconstruction efficiencies, $10^{21}$ protons on target (p.o.t) will be required for this.   
To date MiniBooNE has received 3.66x$10^{20}$ p.o.t.  This produces 
greater than 380K neutrino events, out of $\approx$1 million expected
for the final appearance analysis, as estimated by our event generation model.
The 380K can be broken down in the following way : 250 K 
from charged current (CC) events, 60K from neutral current (NC) elastic, and 26K from NC $\pi^0$ events.  
The charged current results are covered by another speaker in these proceedings.~\cite{Morgan}

\subsection{NC Elastic}
Neutral current elastic events will be used for measuring the component of the proton 
spin carried by strange quarks. This event class is also currently 
being used to study scintillation properties of oil and the low
energy response of the detector which are both additionally
useful for the $\nu_e$ appearance analysis. 
 Current results for this analysis have been presented previously.~\cite{jocelyn}  

\subsection{NC $\pi^0$}
Neutral current $\pi^0$ events are the dominant mis-identification background to the $\nu_e$ appearance 
oscillation analysis.  The cross section combined with the angular production 
distribution also constrains the coherent and resonant production mechanism of these events.  

\begin{figure}[t]
\psfig{file=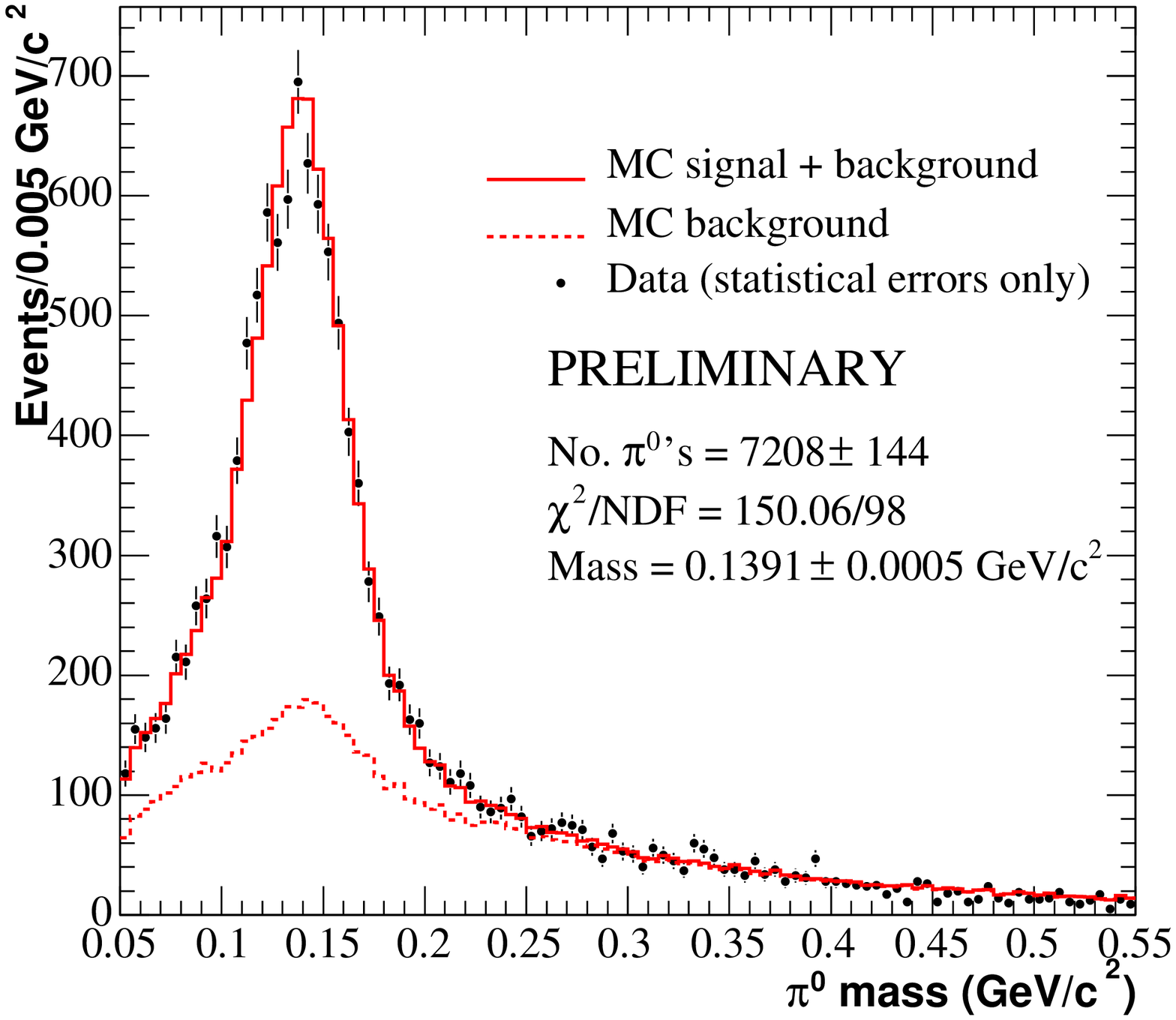, width = 6cm}
\hspace{0.2cm}
\psfig{file=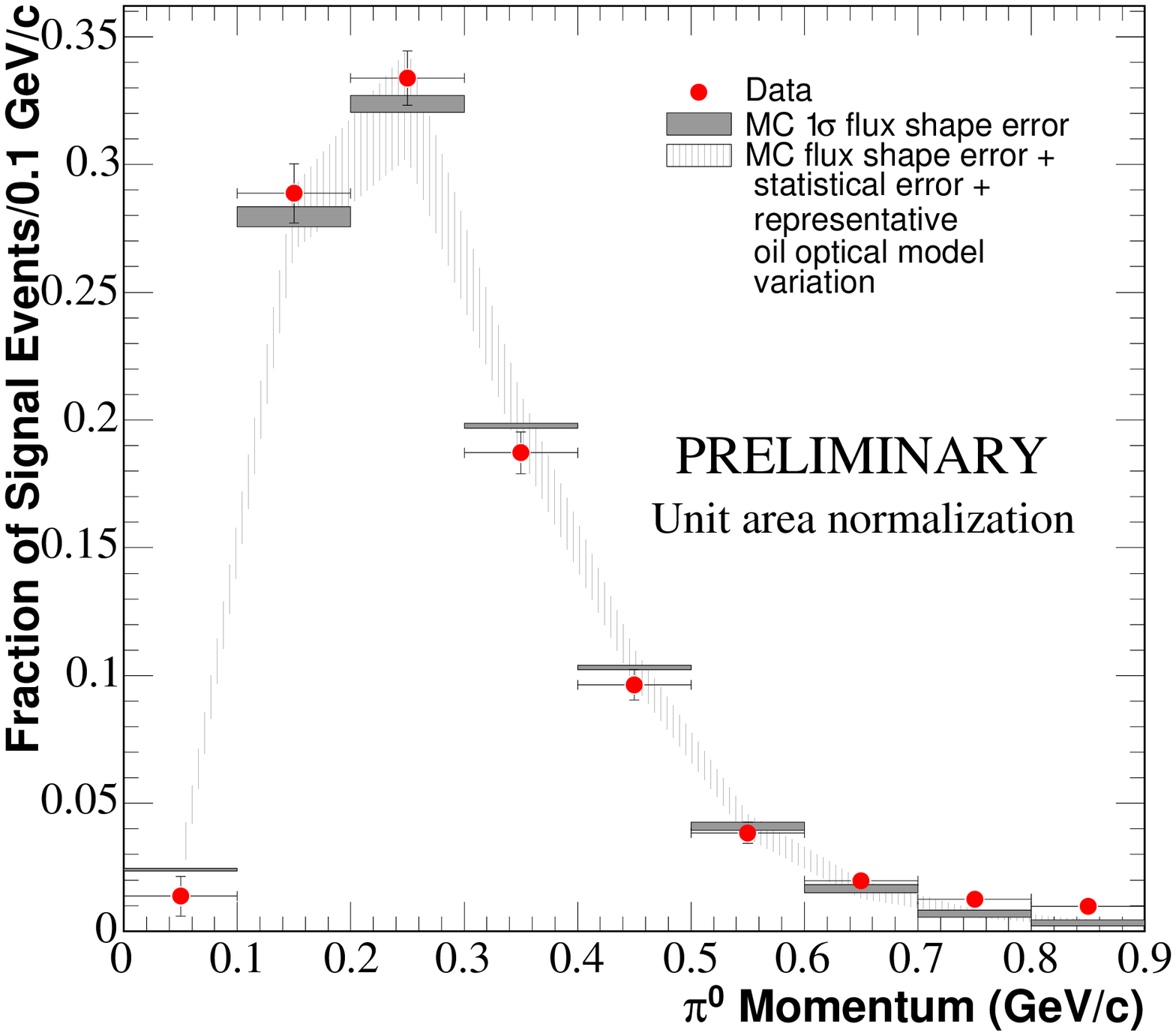, width = 6cm}

\caption{ $\pi^0$ reconstructed mass peak (left), and momentum spectrum (right).  The mass peak signal is from NC resonant and coherent single $\pi^0$ events.  Backgrounds to these events contain interactions which produce $\pi^0$; the peak near the $\pi^0$ mass in the background is expected.  Errors shown are shape errors: dark bands are from flux errors, and the light bands are from the current understanding of the optical model.  \label{fig:one} }
\end{figure}

Reconstruction of the $\pi^0$ mass peak gives a 55\% sample purity with 42\% efficiency 
in the mass peak.  The events in the mass peak are used to plot the $\pi^0$ momentum.
The $\pi^0$ momentum spectrum plot shows excellent agreement between data and Monte Carlo. (See Figure~\ref{fig:one}.) 

\section{Conclusions}

MiniBooNE analyses and data taking are proceeding well.  
Analysis algorithms have successfully identified 
CC, NC $\pi^0$, and NC elastic interactions. We have collected 3.66x$10^{20}$ p.o.t.
and data taking is expected to continue through 2005.



\begin{thebibliography}{0}
\bibitem{MiniBooNE} http://www-boone.fnal.gov
\bibitem{LSND} A. Aguilar et al.  Phys.Rev.D64 (2001) 112007
\bibitem{E910} http://www.phy.bnl.gov/$\sim$e910/html/home.html
\bibitem{HARP} A. Villanueva.  arXiv:hep-ex/040603
\bibitem{Roe04} B. Roe, H. Yang, J. Zhu, Y. Liu, I. Stancu, G. McGregor.  arXiv:physics/0408124
\bibitem{Morgan} M. Wascko, ``Measuring $\nu_{\mu}$ Charged-Current Interactions in MiniBooNE'', these proceedings
\bibitem{jocelyn} J. Monroe.  arXiv:hep-ex/0406048
\end{thebibliography}
\end{document}